\documentclass[12pt]{article}
\usepackage{amssymb}

\begin{document}

\title{Dynamical Mechanism for Varying Light Velocity as a Solution
to Cosmological Problems}

\author{M. A. Clayton\thanks{\texttt{clayton@medb.physics.utoronto.ca}}
and J. W. Moffat\thanks{\texttt{moffat@medb.physics.utoronto.ca}}}

\date{\today}

\maketitle
\begin{center}
\textit{Department of Physics, University of Toronto, Toronto,
Ontario M5S 1A7, Canada}
\end{center}

\begin{abstract}
A dynamical model for varying light velocity in cosmology is
developed, based on the idea that there are two metrics in
spacetime. One metric $g_{\mu\nu}$ describes the standard
gravitational vacuum, and the other ${\hat
g}_{\mu\nu}=g_{\mu\nu}+\beta\psi_\mu\psi_\nu$ describes the
geometry through which matter fields propagate.  Matter
propagating causally with respect to $\hat{g}_{\mu\nu}$ can
provide acausal contributions to the matter stress-energy tensor
in the field equations for $g_{\mu\nu}$, which, as we explicitly
demonstrate with perfect fluid and scalar field matter models,
provides a mechanism for the solution of the horizon, flatness and
magnetic monopole problems in an FRW universe. The field equations
also provide a ``graceful exit'' to the inflationary epoch since
below an energy scale (related to the mass of $\psi_\mu$) we
recover exactly the standard FRW field equations.

\end{abstract}


\textbf{keywords:} cosmology, inflation, alternative gravitational
theories, causality.

\section{Introduction}

The standard inflationary epoch scenario can explain several of
the observed features of the universe such as the flatness and
homogeneity of the present universe, as measured by the cosmic
microwave background (CMB) measurements, the isotropy of the
universe (horizon problem), and the lack of relic magnetic
monopoles~\cite{Guth:1981,Linde:1990,Turner:1995}. Much effort has
gone into constructing viable models of inflation, notably using a
scalar inflaton associated with a large vacuum energy
(cosmological constant) in the early universe. To invoke the
required e-folds of inflation, it is necessary to begin with an
approximately pre-inflationary homogeneous universe, otherwise,
not enough e-folds of inflation can be achieved to solve the
horizon and flatness problems. The generic prediction of
inflationary models is that $\Omega=1$, where
$\Omega=\rho/\rho_{\rm crit}$.

However, because the observed baryonic matter in the universe is
not sufficient for this, one is forced into a scenario where most
of the matter in the universe is non-baryonic dark
matter~\cite{Turner:1995}. Recently, the cosmological constant
$\Lambda$ has been replaced with a dynamical, time dependent and
spatially inhomogeneous component whose equation of state differs
from the standard matter, dark matter and
radiation~\cite{Caldwell+Dave+Steinhardt:1998}. This new
contribution to the cosmological energy density (called
``quintessence" or Q-component) can be described by fundamental
fields or macroscopic objects, such as light cosmic strings. Fits
to recent data are superior to those using $\Lambda$ and cold dark
matter.

In the following, we shall develop a dynamical model of the
superluminary phase transition that can solve the horizon,
flatness and magnetic monopole relic problems, and can furnish a
prediction for the temperature fluctuations observed in the CMB.
It provides a specific dynamical mechanism to explain the origin
of the spontaneous symmetry breaking of local Lorentz and
diffeomorphism invariance postulated in earlier
publications~\cite{Moffat:1993a,Moffat:1993b,Moffat:1998}. This
leads to a concrete model in which light effectively travels at a
much larger speed in the very early universe and undergoes a phase
transition to its standard speed at some critical time
$t=t_{\mathit{pt}}$, when the local Lorentz and diffeomorphism
symmetries are restored.

Albrecht, Magueijo and Barrow have also proposed models of varying
light speed as possible solutions to the initial value problems in
cosmology~\cite{Albrecht+Maguiejo:1999,Barrow:1999}.  The model
proposed here is distinct in that we are not considering the
possibility that what was heretofore considered a `constant' of
nature is time varying.  Instead we begin by motivating the type
of theory that can lead to the physical idea of a speed of light
which is `dynamical', and, after restricting ourselves to a
concrete realization of this, only then proceed to show that the
standard cosmological problems can be solved in the assumption of
homogeneity and isotropy.  This is certainly closer in spirit to
the philosophy that motivated the development of general
relativity, and therefore we feel that this is a significant
(albeit philosophical) step forward. On a more practical side,
although we also expect to see preferred-frame effects locally,
there are no non-dynamical fields in our diffeomorphism-invariant
formulation and therefore these effects are the result of local
dynamics rather than a global frame chosen at the outset.  We
therefore expect that our model will be less tightly constrained
by experiment~\cite{Will:1998}.

One of the important improvements achieved by the new scenario
described here, is that the initial conditions of the universe are
not as restrictive as those required by the inflationary model.
Moreover, although we solve the field equations of the theory
assuming that the universe is initially a
Friedmann-Robertson-Walker (FRW) flat and homogeneous universe, we
expect that the initial universe can be generalized to more
complicated inhomogeneous models without losing the generic
solutions to the cosmological problems.

\section{Bimetrics and Field Equations}

The idea here is to present a model that embodies the physical
content of a ``varying speed of light'' in a diffeomorphism
invariant manner--without introducing a global preferred reference
frame into spacetime. To accomplish this, we note that the causal
propagation of electromagnetic fields is determined from the
spacetime metric that appears in Maxwell's equations, and
therefore changing the speed at which light propagates is
accomplished by making (non-conformal) alterations to this metric.
In order for this to have physical consequences we need to ensure
that this is not the only metric in spacetime, so that we can
therefore concretely talk about the speed of light as being
different than the speed of propagation of other fields. (This
differs from Brans-Dicke theory in a fundamental way: there the
scalar field is essentially a conformal factor and the light cones
derived in the ``Einstein frame'' and the ``Jordan frame'' are
identical, whereas here we are working with two metrics with
inequivalent causal structure.)

As a simple and functional model, we consider the introduction of
a covector field $\psi_\mu$ which relates the ``gravitational
metric'' $g_{\mu\nu}$ to the ``matter metric'' $\hat{g}_{\mu\nu}$
by
\begin{equation}\label{eq:metrics}
{\hat g}_{\mu\nu}=g_{\mu\nu}+\beta\psi_\mu\psi_\nu,
\end{equation}
where $\beta>0$ is a dimensionless constant.  The class of models
we consider is described by the action
\begin{equation}\label{eq:total action}
S_{\rm tot}=S_{\rm gr}[g]+S_\psi[\psi,g] +S_{\rm
matter}[\hat{g},{\rm matter\, fields}],
\end{equation}
where
\begin{equation}
\label{eq:GR action}
  S_{\rm gr}[g]=-\frac{1}{\kappa}\int dtd^3x\sqrt{-g}(R[g]-2\Lambda),
\end{equation}
is the usual Einstein-Hilbert action, $\kappa=16\pi G/c^4$ and
$\Lambda$ is the cosmological constant. Mindful of the issues
involved in constructing well-behaved vector field
actions~\cite{Isenberg+Nester:1977} and noting that the structure~(\ref{eq:metrics})
is not invariant under local $\mathrm{U}(1)$ transformations, we
assume a Maxwell-Proca action for the covector field ($m=\mu
c/\hbar$ has dimensions of an inverse length):
\begin{equation}
S_{\psi}[\psi,g]=\frac{1}{\kappa}\int
dtd^3x\sqrt{-g}\Bigl(-\frac{1}{4}B^2+\frac{1}{2}m^2\psi^2\Bigr),
\end{equation}
where $B_{\mu\nu}:=\partial_\mu\psi_\nu-\partial_\nu\psi_\mu$,
$B^2:=g^{\mu\nu}g^{\alpha\beta}B_{\mu\alpha}B_{\nu\beta}$ and
$\psi^2:=g^{\mu\nu}\psi_\mu\psi_\nu$.  We assume that the matter
field action is one of the standard forms, but constructed out of
$\hat{g}_{\mu\nu}$, and therefore the field equations guarantee
that the conservation laws $\hat{\nabla}_\nu
T^{\mu\nu}_{\mathrm{matter}}[\hat{g}]=0$, where $\hat{\nabla}_\nu$
denotes the covariant derivative with respect to the
$\hat{g}_{\mu\nu}$ metric connection, and
\begin{equation}
T^{\mu\nu}_{\mathrm{matter}}[\hat{g}]=
\frac{2}{\sqrt{-\hat{g}}}\hat{g}^{\mu\alpha}\hat{g}^{\nu\beta}
\biggl(\frac{\delta S_{\mathrm{matter}}[\hat{g}]}
{\delta\hat{g}^{\alpha\beta}}\biggr),
\end{equation}
are satisfied.

 Variation of~(\ref{eq:total action}) with respect
to $g_{\mu\nu}$ and $\psi_\mu$ leads to the field equations:
\begin{equation}
\label{eq:GR FEQ} \sqrt{-g}(G^{\mu\nu}[g]-\Lambda
g^{\mu\nu})=\frac{1}{2} \sqrt{-g}T^{\mu\nu}[g,\psi]
+\frac{\kappa}{2}\sqrt{-\hat{g}}T^{\mu\nu}_{\rm
matter}[\hat{g}],\\
\end{equation}
\begin{equation}
\label{eq:covector FEQ} \sqrt{-g}\Bigl(-\nabla_\nu
B^{\mu\nu}+m^2\psi^\mu \Bigr) =\beta\kappa
\sqrt{-\hat{g}}T^{\mu\nu}_{\rm matter}[\hat{g}]\psi_\nu,
\end{equation}
where $\nabla_\nu$ denotes the covariant derivative formed from
the $g_{\mu\nu}$ metric connection, and
\begin{equation}
T_{\mu\nu}=-B_{\mu\alpha}{B_\nu}^\alpha+\frac{1}{4}g_{\mu\nu}B^2
  +m^2\psi_\mu \psi_\nu -\frac{1}{2}g_{\mu\nu}
  m^2\psi^2.
\end{equation}
It is a straightforward exercise to show that the field equations
and matter conservation laws are consistent with the Bianchi
identities.  (Note that this type of ``vector-tensor'' theory is
distinct from those considered, for example, in~\cite{Will:1993}.)

Some comments on this construction are in order.  Note that the
gravitational metric fields and the covector $\psi_\mu$ propagate
on the geometry described by $g_{\mu\nu}$, whereas all other
matter fields will propagate on the geometry described by
$\hat{g}_{\mu\nu}$.  Thus if we consider the motion of a
(non-gravitational) test particle, it is reasonable to assume that
it is the geodesics of $\hat{g}_{\mu\nu}$ that are of physical
interest.  It is also very important to recognize that the energy
conditions normally imposed on the matter stress-energy tensor no
longer have the same implications for the gravitational field
equations.  To illustrate this, consider a vector field $v^\mu$
which is null with respect to the matter metric: $\hat{g}(v,v)=0$.
From~(\ref{eq:metrics}) we find that $g(v,v)=-\beta(\psi_\mu
v^\mu)^2\le 0$, and therefore $v^\mu$ may be spacelike \textit{or}
null with respect to $g_{\mu\nu}$ (which motivates the choice
$\beta>0$). This will manifest itself in Section~\ref{sect:HI
model} as a fluid that behaves in a perfectly causal way with
respect to the matter metric ${\hat g}_{\mu\nu}$, but appears as
an acausally, propagating fluid in~(\ref{eq:GR FEQ}).

The field equations~(\ref{eq:covector FEQ}) have the important
property that $\psi_\mu=0$ is always a solution regardless of the
matter content of spacetime, in which case the conventional
general relativity coupled to matter models are realized and there
is no conflict with experiment. In regions of spacetime where
$\psi_\mu$ is nonvanishing and $g(\psi,\psi)>0$, we can restrict
ourselves to frames that are aligned with the covector field:
$\psi_\mu\rightarrow(1,0,0,0)$, and we have reduced the gauge
group of the orthonormal frames to $\mathrm{O}(3)$. Thus we see
that $\psi_\mu\neq 0$ plays the role of a vacuum condensate
$\langle\psi_\mu\rangle_0$ that can be said to spontaneously
`break' local Lorentz invariance.

Note that the model that we have introduced here is a ``metric
theory of gravity''~\cite{Will:1998} in the sense that all matter
and non-gravitational fields respond to the metric
$\hat{g}_{\mu\nu}$. The dynamics that determine $\hat{g}_{\mu\nu}$
involve the tensor $g_{\mu\nu}$ as well as the covector $\psi_\mu$
and therefore preferred frame effects are possible. However, since
we expect that the vector field will essentially lead to repulsive
effects (the presence of $\psi_\mu$ locally increases the speed of
matter propagation, thereby effectively decreasing the
gravitational coupling to the matter) and the magnitude of the
vector field is dependent upon the local matter energy density, we
expect that the effects of $\psi_\mu$ will be negligible in the
present universe.

Most of what appears in this work could be reproduced using a
scalar field to define the matter metric as (for example)
$\hat{g}_{\mu\nu}=g_{\mu\nu}+\beta\nabla_\mu\phi\nabla_\nu\phi$.
Although in some ways the scalar field driven mechanism is
preferable, the model presented here is much cleaner conceptually
as well as algebraically.  We will return to the scalar field
driven case in a later publication.

\section{A Homogeneous and Isotropic Model}
\label{sect:HI model}

We now examine what effect this additional structure has on the
standard cosmological scenario.  Beginning with a one-parameter
family of perfect fluid matter sources ($p\propto \rho$), we find
a solution that demonstrates that while the matter energy density
is greater than a threshold $\rho_{pt}$ (to be identified later),
the matter will experience an inflating universe regardless of the
equation of state.  These matter models do not provide enough
e-folds to solve the Horizon problem unless $p\approx -c^2\rho$,
which is precisely the relation one derives from a ``slowly
rolling'' scalar field in inflationary scenarios.  We show that
the mechanism that we are proposing enhances even the simplest
inflationary scalar field model to the point where no fine-tuning
is necessary.

Assuming that all the fields in the theory are spatially
homogeneous and isotropic leads us to the FRW form of the
gravitational metric in comoving coordinates:
\begin{equation}
\label{eq:gravitational metric}
g_{\mu\nu} dx^\mu  dx^\nu
=c^2dt^2-R^2(t)\biggl[\frac{dr^2}{1-kr^2}+r^2d\theta^2+r^2\sin^2\theta
d\phi^2\biggr],
\end{equation}
where we will be employing a dimensionless radial coordinate $r$
and $k=0,\pm 1$ for the flat, closed and hyperbolic spatial
topologies.  Note that this has fixed the time reparameterization
invariance of the theory, and because the spacetime symmetries
require that $\psi_\mu=(c\psi_0(t),0,0,0)$, the matter metric is
given by
\begin{equation}
\label{eq:matter metric}
\hat{g}_{\mu\nu} dx^\mu  dx^\nu=
c^2dt^2[1+\beta\psi_0^2(t)]-R^2(t)\biggl[\frac{dr^2}{1-kr^2}+r^2d\theta^2
+r^2\sin^2\theta d\phi^2\biggr].
\end{equation}
From these we also find that
\begin{equation}
\sqrt{-\hat{g}} =\bigl(1+\beta \psi_0^2(t)\bigr)^{1/2}\sqrt{-g}.
\end{equation}

As a simple matter model, we consider a perfect fluid
\begin{equation}
\label{eq:perfect Fluid}
 T^{\mu\nu}_{\rm matter}=(\rho+\frac{p}{c^2})u^\mu u^\nu -p\hat{g}^{\mu\nu},
\end{equation}
where the vector field is normalized as $\hat{g}_{\mu\nu}u^\mu
u^\nu=c^2$, resulting in
\begin{equation}
\label{eq:normalization of vector}
u^0=1/\sqrt{1+\beta\psi_0^2(t)}.
\end{equation}
The matter conservation equations lead to
\begin{equation}\label{eq:matter conservation}
{\dot\rho}(t)+3\biggl(\rho(t)+\frac{p(t)}{c^2}\biggr)\biggl(\frac{{\dot
R}(t)}{R(t)}\biggr)=0,
\end{equation}
and the single nontrivial field equation derived from
(\ref{eq:covector FEQ}) is equivalent to
\begin{equation}
\label{eq:vector reduced}
 \psi_0(t)\biggl[m^2\sqrt{1+\beta\psi_0^2(t)}
-\beta\kappa c^2 \rho(t)\biggr]=0.
\end{equation}

We will examine the nontrivial solution
\begin{equation}\label{eq:nontrivial}
\beta \psi_0^2(t)=(\rho(t)/\rho_{\mathit{pt}})^2 -1,
\end{equation}
which we will refer to as the ``broken phase" in analogy with
spontaneous symmetry breaking. We have identified the time at
which $\psi_0(t)$ vanishes as $t_{\mathit{pt}}$ and for
convenience defined
\begin{equation}
\rho_{\mathit{pt}}: =m^2/(\beta\kappa c^2),\quad
H_{\mathit{pt}}:=\sqrt{c^2m^2/(6\beta)},
\end{equation}
the latter of which we will identify later as (approximately) the
Hubble function evaluated at $t_{\mathit{pt}}$.  Since we expect
that $\rho(t)$ will decrease from the initial singularity, if we
assume that the universe begins in the ``broken phase'', then
$\psi_0(t)$ will decrease to zero.  After this time it will be
forced to vanish identically because the ``broken phase'' is
unavailable to the system when $\rho(t)<\rho_{\mathit{pt}}$, and
so for $t>t_{\mathit{pt}}$ we smoothly match the solution to that
of a standard FRW cosmological model~\cite{Kolb+Turner:1990}.

Using~(\ref{eq:nontrivial}), the nontrivial Friedmann equation
obtained from~(\ref{eq:GR FEQ}) becomes
\begin{equation}
\label{eq:FRW symmetric} \frac{{\dot
R}^2(t)}{R^2(t)}+\frac{kc^2}{R^2(t)}=
\frac{1}{3}c^2\Lambda+\frac{1}{2}H_{\mathit{pt}}^2
\Bigl[1+\Bigl(\frac{\rho(t)}{\rho_{\mathit{pt}}}\Bigr)^2\Bigr].
\end{equation}
Adopting the one-parameter family of equations of state
($\omega>0$ is a fixed constant)
\begin{equation}
p(t)=(\omega/3-1)c^2\rho(t),
\end{equation}
we use~(\ref{eq:matter conservation}) to find that
\begin{equation}\label{eq:density}
\rho(t)=\rho_{\mathit{pt}}\biggl(\frac{R_{\mathit{pt}}}{R(t)}\biggr)^\omega,
\end{equation}
where we also define $R_{\mathit{pt}}:=R(t_{\mathit{pt}})$. This
is of the same form as the standard Friedmann equation with an
effective cosmological constant given by
$\Lambda_{\mathit{eff}}=\Lambda+m^2/(4\beta)$, an effective energy
density $\rho_{\mathit{eff}}=\rho^2(t)/(2\rho_{\mathit{pt}})$, and
an effective pressure $p_{\mathit{eff}}=5c^2\rho_{\mathit{eff}}/3$
that appears to violate the causal energy requirements. This is
not unexpected, for the fluid will appear to violate causality in
the ``gravitational frame'', while being perfectly causal in the
``matter frame''.  Although it is relatively trivial to include
it, we will only consider solutions with vanishing bare
cosmological constant ($\Lambda=0$) here.

The equation~(\ref{eq:FRW symmetric}) is difficult to solve in
general, however, if $\omega\ge 1$ and we require that
\begin{equation}\label{eq:k approx}
H_{\mathit{pt}}^2R_{\mathit{pt}}^2/c^2\gg 1,
\end{equation}
which is essentially the condition that the ``size'' of the
universe at $t_{\mathit{pt}}$ is much larger than the fundamental
length scale of $\psi_\mu$, then it is straightforward to show
that the effect of $k$ in the Friedmann equation is negligible
during this phase.  (This is also true for $0<\omega<1$ after the
length scale of the universe has had time to grow much larger than
$m^{-1}$.) In this approximation (or assuming that $k=0$) we find
the solution
\begin{equation}
\label{sinh}
 R(t)=R_{\mathit{pt}}
 \sinh^{1/\omega}\bigl[(\omega H_{\mathit{pt}}/\sqrt{2})(t-t_{\mathit{pt}})
 +\mathrm{arcsinh}(1)\bigr].
\end{equation}
Identifying the initial singularity at time $t=t_{init}$ by
$R(t_{init})=0$, we observe that the universe remains in this
phase for a time $t_{pt}-t_{init}=\sqrt{2}\mathrm{arcsinh}(1)/
(\omega H_{\mathit{pt}})$. The horizon scale on this interval:
\begin{equation}\label{eq:gravitational horizon scale}
d_H(t_{\mathit{init}};t_{\mathit{pt}})=\frac{c(t_{\mathit{init}}-t_{\mathit{pt}})}
{\mathrm{arcsinh}(1)}
\int_0^{\mathrm{arcsinh}(1)}\frac{dy}{\sinh^{1/\omega}(y)},
\end{equation}
is finite for $\omega>1$ (and, indeed, if $\omega=3$ or $\omega=4$
does not differ significantly from the usual radiation or matter
dominated result: $2c(t_{\mathit{init}}-t_{\mathit{pt}})$) and
diverges for $0<\omega\le 1$.

The matter metric~(\ref{eq:matter metric}) has
$\hat{g}_{00}=c^2(\rho(t)/\rho_{\mathit{pt}})^{2}$, and so
$\hat{g}_{\mu\nu}$ may be put into a comoving frame by introducing
the coordinate
$d\tau=dt(\rho(t)/\rho_{\mathit{pt}})=dt(R_{\mathit{pt}}/R(t))^\omega$.
Requiring that $\tau_{\mathit{pt}}=t_{\mathit{pt}}$, we find
\begin{equation}\label{eq:t tau}
  \tau=t_{\mathit{pt}}+\frac{\sqrt{2}}{\omega H_{\mathit{pt}}}
  \ln\biggl\{\frac{\tanh[((\omega H_{\mathit{pt}}/\sqrt{2})(t-t_{\mathit{pt}})+\mathrm{arcsinh}(1))/2]}
  {\tanh(\mathrm{arcsinh}(1)/2)}\bigg\},
\end{equation}
from which it is clear that the finite coordinate time between the
initial singularity and $t_{\mathit{pt}}$ in the gravitational
frame is mapped into the infinite coordinate interval
$\tau\in(-\infty,t_{\mathit{pt}}]$ in the matter frame. Using
these results, we find
\begin{eqnarray}
\label{eq:Rtau}
  R(\tau)=R_{\mathit{pt}}\sinh^{-1/\omega}(x),\quad
  \rho(\tau)=\rho_{\mathit{pt}}\sinh(x),\quad
  H(\tau)=(H_{\mathit{pt}}/\sqrt{2})\coth(x),
\end{eqnarray}
where for convenience we have defined
\begin{equation}\label{eq:x definition}
x:=(\omega
H_{\mathit{pt}}/\sqrt{2})(\tau_{\mathit{pt}}-\tau)+\mathrm{arcsinh}(1),\quad
x\in[\mathrm{arcsinh}(1),\infty).
\end{equation}
Note that near the $R=0$ singularity ($\tau\ll
\tau_{\mathit{pt}}$), we have the approximate form
\begin{equation}\label{eq:R approx}
R(\tau)\approx
2^{1/\omega}R_{\mathit{pt}}\exp(H_{\mathit{pt}}\tau/\sqrt{2}),
\end{equation}
and we see inflationary behaviour~\cite{Guth:1981} irrespective of
the value of $\omega$.  These results may also be derived directly
by re-writing~(\ref{eq:FRW symmetric}) in terms of $\tau$ to find
\begin{equation}\label{eq:Friedmann tau}
\frac{{\dot R}^2(\tau)}{R^2(\tau)}+\frac{kc^2}{R^2_{\mathit{pt}}}
\Bigl(\frac{R(\tau)}{R_{\mathit{pt}}}\Bigr)^{2(\omega-1)}=
\frac{1}{3}c^2\Lambda\Bigl(\frac{R(\tau)}{R_{\mathit{pt}}}\Bigr)^{2\omega}
+\frac{1}{2}H_{\mathit{pt}}^2
\Bigl[1+\Bigl(\frac{R(\tau)}{R_{\mathit{pt}}}\Bigr)^{2\omega}\Bigr],
\end{equation}
and setting $\Lambda=0=k$.

Evaluating the horizon scale on $(\tau,\tau_{\mathit{pt}})$ we
find
\begin{equation}\label{eq:matter horizon scale}
d_H(\tau;\tau_{\mathit{pt}})=
2c(\tau_{\mathit{pt}}-\tau)\Biggl[\frac{1}{2(x-\mathrm{arcsinh}(1))}
\int_{\mathrm{arcsinh}(1)}^x \sinh^{1/\omega}(y)dy\Biggr],
\end{equation}
which diverges as $\tau\rightarrow -\infty$ for any $\omega>0$,
and therefore it is possible to solve the horizon
problem~\cite{Hu+Turner+Weinberg:1994}. We have written it in this
form to emphasize that this divergence is not solely due to the
fact that the time interval is infinite in the limit; the part
of~(\ref{eq:matter horizon scale}) in square brackets diverges
separately. Of course, the horizon scales~(\ref{eq:gravitational
horizon scale}) and~(\ref{eq:matter horizon scale}) have different
physical meanings.  The first describes the proper size of regions
that are causally connected via gravitational radiation at
$\tau_{\mathit{pt}}$, whereas the latter describes the size of the
regions that are connected causally by the propagation of matter
fields.  Although it may be that solving the horizon problem in
the matter sector is sufficient, if $0<\omega\le 1$ then the
gravitational horizon scale is also ``inflated''. We should
emphasize that although it is precisely this case that drives the
conventional inflationary models, what we have here is more like
an ``enhanced inflation''--assuming that $\psi_0\neq 0$ initially,
\textit{any} form of matter energy will cause the universe to
inflate. As we shall see, the inflationary effect of matter with
negative pressure is enhanced over conventional models of
inflation.

It is perhaps slightly disconcerting that the universe as it
appears in the matter frame is infinitely old. Our gravitational
model, however, is classical, and we do not expect it to be
accurate when matter energy densities become greater than the
Planck density $\rho_{\mathit{P}}:=c^5/(\hbar G^2)\approx
5.2\times 10^{93}g/\mathit{cm}^3$. From~(\ref{eq:Rtau}) this
occurs at a time $\tau_{qg}$ defined by
$\sinh(x_{qg})\approx\rho_{\mathit{P}}/\rho_{\mathit{pt}}$, and
therefore between $\tau_{\mathit{qg}}$ and $\tau_{\mathit{pt}}$
the radial scale of the universe increases by a factor
\begin{equation}\label{eq:rho Planck relation}
(\rho_{\mathit{P}}/\rho_{\mathit{pt}})^{1/\omega}= :e^N,
\end{equation}
where $N\gtrsim 60$ to solve the horizon
problem~\cite{Turner:1995}. Using this to write
$H_{\mathit{pt}}^{-1}\approx
(8\pi/3)^{-1/2}t_{\mathit{P}}\exp(\omega N/2)$, where
$t_{\mathit{P}}:=(G\hbar/c^5)^{1/2}\approx 5.4\times 10^{-44}
\mathit{sec}$ is the Planck time, it is straightforward to
determine the coordinate time that the universe spends in this
phase:
\begin{equation}\label{eq:phase time}
\tau_{\mathit{pt}}-\tau_{\mathit{qg}} \approx
A(N\omega)NH_{\mathit{pt}}^{-1} \approx A(N\omega)N
t_{\mathit{P}}\sqrt{3/(8\pi)}\exp(\omega N/2),
\end{equation}
where $A(N
\omega):=\sqrt{2}(\mathrm{arcsinh}(\exp(N\omega))-\mathrm{arcsinh}(1))/(N\omega)\in
(1,\sqrt{2})$. We see, therefore, that the time spent in this
phase of the universe is longer than would be possible in
conventional models by a scaling factor $\approx N$.

As one would expect from~(\ref{eq:FRW symmetric})
or~(\ref{eq:Friedmann tau}), if $\omega<1$ we have a solution to
the flatness problem that mimics that which is provided by
inflation (although admittedly, the status of the flatness problem
is not completely clear~\cite{Cho+Kantowski:1995}). Evaluating
$\Omega-1$ and comparing it to the value that obtains at
$t_{\mathit{pt}}$ we find:
\begin{equation}\label{eq:one}
\vert\Omega-1\vert=\vert\Omega-1\vert_{t_\mathit{pt}}
\frac{2c^2\mathrm{tanh}^2(y)}{\mathrm{sinh}^{2/\omega}(y)},
\end{equation}
where $y=(\omega
H_{\mathit{pt}}/\sqrt{2})(t-t_{\mathit{pt}})+\mathrm{arcsinh(1)}$.
As $t\rightarrow t_{\mathit{init}}$ ($\tau\rightarrow
\tau_{\mathit{qg}}$) we find that $\Omega-1$ vanishes for
$\omega>1$ and diverges for $\omega <1$, and therefore the extreme
fine-tuning that is necessary in non-inflationary models is
avoided~\cite{Guth:1981,Hu+Turner+Weinberg:1994}. For $\omega<1$
models the solution to the magnetic monopole problem is identical
to that described by Guth~\cite{Guth:1981} and will not be
repeated here.

It is noteworthy that models with $\omega\approx 0$ achieve a
given inflation factor in the shortest possible time. Although as
$\omega N\rightarrow 0$ the time interval~(\ref{eq:phase time})
becomes very small:
$\tau_{\mathit{pt}}-\tau_{\mathit{qg}}\rightarrow
N(8\pi/3)^{-1/2}t_{\mathit{P}}$, it is not difficult to show that
the horizon scale
$d_H(\tau_{\mathit{qg}};\tau_{\mathit{pt}})\rightarrow
2c(\tau_{\mathit{pt}}-\tau_{\mathit{qg}}) (\exp(N)-1)/(2N)$. It is
also interesting to note that within the limits of our
approximation, if the broken phase ends near electroweak symmetry
breaking: $1/(2H_{\mathit{pt}})\approx 10^{-11}\mathit{sec}$, then
we are led to $N\omega\approx 152$. If we assume that radiative
energy dominates the universe back to $\tau_{\mathit{qg}}$ then we
get $N\approx 38$ which is not quite sufficient to solve the
horizon problem.  Requiring that $N\approx 60$ would imply that
the broken phase lasts well into the observable universe. Choosing
instead the broken phase to end at $1/(2H_{\mathit{pt}})\approx
10^{-35}\mathit{sec}$ (roughly corresponding to the temperature at
which some GUT symmetry is broken
spontaneously~\cite{Linde:1990}), we find $N\omega\approx 42$,
clearly requiring that $\omega<1$.

To achieve this with a more realistic matter model, we adopt a
scenario familiar from inflation, namely, we assume that the
universe exits the quantum gravitational stage with the matter
energy of the universe dominated by a Higgs field close to a
``false vacuum''. Assuming the ``slow roll'' approximation where
the kinetic energy of the scalar field is (at least initially)
dominated by the potential energy, the effective pressure is
negative and the effective equation of state has $\omega\approx
0$.

Working in the gravitational frame, the field equation for the
scalar field
\begin{equation}
\ddot{\phi}(t)+3H(t)\dot{\phi}(t)+c^2\delta V[\phi]=0,
\end{equation}
assuming that the acceleration term can be neglected and that
$H(t)\approx H_\phi$, has the approximate solution
\begin{equation}\label{eq:scalar approx}
\phi(t)\approx\phi_0\exp\Bigl[\frac{c^2m_\phi^2}{6 H_\phi}
(t-t_{\mathit{qg}})\Bigr].
\end{equation}
For clarity we use the minimal Higgs potential (we have chosen
$\phi$ to be dimensionless)
$V[\phi]=\lambda(\phi^2-m_\phi^2l_{\mathit{P}}^2/(2\lambda))^2/(4l_{\mathit{P}}^2)$,
where $\hbar m_\phi/c$ is the mass of the scalar field in the
physical vacuum which we will assume corresponds to the GUT scale
$\approx 5\times 10^{14}\mathit{GeV}$, $\lambda$ is a
dimensionless coupling constant and
$l_{\mathit{P}}:=(G\hbar/c^3)^{-1/2}\approx 1.3\times
10^{-33}\mathit{cm}$ is the Planck length. To find~(\ref{eq:scalar
approx}) we have also made the approximation $\delta
V[\phi]\approx -m_\phi^2\phi/2$; note that~(\ref{eq:scalar
approx}) is valid provided that $H_\phi^2 \gg c^2 m^2_\phi/18$.

The energy density of a homogeneous and isotropic scalar field
$\rho=(\frac{1}{2}\dot{\phi}^2+c^2V[\phi])/(\kappa c^4)$ enters
into the Friedmann equations~(\ref{eq:FRW symmetric}) proportional
to $\rho^2(t)$, and using $\rho(t)\approx l_{\mathit{P}}^2
m_\phi^4/(16\lambda\kappa c^2)$, we find that
\begin{equation}\label{eq:Rt approximation}
\frac{\dot{R}^2(t)}{R^2(t)}\approx
H_\phi^2:=\frac{1}{2}H_{\mathit{pt}}^2 \Bigl[1+\Bigl(\frac{\beta
l_{\mathit{P}}^2 m_\phi^4}{16\lambda m^2}\Bigr)^2\Bigr].
\end{equation}
This should be compared to $H_\phi^2\approx c^2 l_{\mathit{P}}^2
m_\phi^4/(96\lambda)$ for a scalar field in the usual inflationary
scenario, leading to the condition $\lambda\ll 3 l_{\mathit{P}}^2
m_\phi^2/16\approx 2.8\times 10^{-10}$ for the slow-roll
approximation to hold, indicating that the scalar field must be
very weakly coupled. This fine-tuning is easily avoided in our
model.  For example, if we assume that $\lambda\approx 1$ and
$m\approx m_\phi$, then $H_\phi^2\approx c^2 m_\phi^2/(12\beta)$,
and we find that our approximation is good provided that $\beta\ll
3/2$.

Since $d\tau/dt\approx \rho(t)/\rho_{\mathit{pt}}\approx \beta
l_{\mathit{P}}^2 m^4_\phi/(16\lambda m^2)$, we see that we have
effectively scaled the speed of light by a constant factor
$c\rightarrow c\beta l_{\mathit{P}}^2 m^4_\phi/(16 \lambda m^2)$,
recovering the scenario originally introduced by
Moffat~\cite{Moffat:1993a}. (Note that for $m\approx m_\phi$ this
indicates that the speed of propagation of matter fields is
enhanced by a factor $\approx 10^{10}\beta/\lambda$ over that of
gravitational fields, which for $\lambda\approx 1 \approx \beta$
is considerably smaller than the value $\approx 10^{30}$ assumed
in~\cite{Moffat:1993a,Moffat:1998}. We shall see though that we
can still get enough $e$-folds without undue fine-tuning.)
Choosing $\tau_{\mathit{qg}}=t_{\mathit{qg}}$, in the matter frame
we find
\begin{equation}
\frac{\dot{R}^2(\tau)}{R^2(\tau)}\approx
\Bigl(\frac{16\lambda m^2}{\beta l_{\mathit{P}}^2m_\phi^4}\Bigr)^2H_\phi^2\rightarrow
R(\tau)=R_{\mathit{qg}}\exp\Bigl[\frac{16\lambda m^2H_\phi}{\beta l_{\mathit{P}}^2m_\phi^4}
(\tau-\tau_{\mathit{qg}})\Bigr],
\end{equation}
and for the scalar field
\begin{equation}
\phi(\tau)\approx\phi_0\exp\Bigl[\frac{8\lambda c^2 m^2} {3\beta
l_{\mathit{P}}^2 m_\phi^2 H_\phi} (\tau-\tau_{\mathit{qg}})\Bigr].
\end{equation}

Defining the number of $e$-folds that the universe expands within
the slow roll approximation (which ends at $\tau_{\mathit{sr}}$)
as
\begin{equation}
N_\phi:=\frac{16\lambda m^2H_\phi}{\beta l_{\mathit{P}}^2 m_\phi^4}
(\tau_{\mathit{sr}}-\tau_{\mathit{qg}}),
\end{equation}
we obtain
\begin{equation}
\phi(\tau_{\mathit{sr}})\approx\phi_0\exp[N_\phi
c^2m_\phi^2/(6H_\phi^2)].
\end{equation}
Choosing $m\approx m_\phi$ this becomes $\phi_0\exp(2\beta
N_\phi)$, and taking $\beta\approx 10^{-3}$ and $N_\phi\approx
100$ we see that the scalar field has not evolved appreciably over
the interval. The requirement that the approximation of the
potential leading to~(\ref{eq:Rt approximation}) is valid
throughout this period is
\begin{equation}\label{eq:phi0 condition}
\frac{4\lambda \phi_0^2}{l_{\mathit{P}}^2
m_\phi^2}\exp\Bigl(\frac{N_\phi c^2 m_\phi^2}{3H_\phi^2}\Bigr)\ll
1,
\end{equation}
which becomes $\phi_0^2\ll l_{\mathit{P}}^2 m_\phi^2/(4\lambda)$
with the above choices.  Assuming that $\lambda\approx 1$ leads to
$\phi_0\lesssim 2\times 10^{-5}$.

We should stress that the ``slow roll'' approximation and the
simple potential model have been introduced in order to show how
to solve the horizon problem in this model without undue
fine-tuning. Since the role of the scalar field is to enhance the
inflationary effect, choosing $1/(2H_{\mathit{pt}})$ anywhere
prior to $\approx 0.01\mathit{sec}$, we expect that there are many
other models with sufficient $e$-folds.  Clearly other scenarios
(inflation at the electroweak scale~\cite{Knox+Turner:1993}, or
possibly more than one scale~\cite{Silk+Turner:1987}) are also
possible.

\section{Concluding Remarks}

A bimetric theory of gravitation is proposed in which two metrics
are associated with spacetime.  The trajectories of test particles
are geodesics of the matter metric ${\hat
g}_{\mu\nu}=g_{\mu\nu}+\beta\psi_\mu\psi_\nu$, the null cones of
which are contained within the light cones of the gravitational
metric $g_{\mu\nu}$. Thus, the gravitational matter spacetime acts
as a ``digravitational" medium and determines the speeds of clocks
associated with the matter.  In the gravitational frame, we have
the standard Einstein equations with matter that can appear to
violate causality, for the matter seems to propagate faster than
the speed of light as defined by the gravitational metric
$g_{\mu\nu}$. But in the matter frame causality is not
violated--matter and gravity propagate less than or at the speed
of light as defined by the metric ${\hat g}_{\mu\nu}$.

The Einstein-matter field equations and the proposed Proca-like
dynamical field equations for $\psi_\mu$ reduce to the equations
of GR when the vector field $\psi_\mu=0$.  We explicitly derived
an exact solution for an FRW universe with $k=\Lambda=0$, based on
an equation of state $p=c^2(\omega/3-1)\rho$. Prior to a time
$t_{\mathit{pt}}$ (which is when the matter density drops below a
threshold defined by the mass of $\psi_{\mu}$) we find that there
are two possible solutions to the field equations. One solution
associated with what we call the ``broken phase" leads to an
inflationary epoch, characterized by an expansion of the universe
with enough inflation to solve the horizon problem, the relic
magnetic monopole problem and the flatness problem.  Following the
phase transition at $t_{\mathit{pt}}$ is the ``unbroken phase'':
only the solution $\psi_\mu=0$ is available to the system. Since
we found that the effect is maximized for $p\approx -c^2 \rho$, we
also considered a scalar field model with a potential that can
lead to inflation ($V\propto\phi^4$).  We found that our mechanism
removed the need for the extreme fine-tuning of the coupling
constant that is required by ordinary inflation.

It is possible to speculate that the phase transition at
$t=t_{pt}$ could occur late enough in the evolution of the
universe (or in patches thereof) that measurable effects could be
observed in the present. This is a possible scenario which we plan
to investigate elsewhere. Finally, we can obtain predictions for
galaxy seeds and CMB temperature fluctuations from the
Maxwell-Proca field equations for $\psi_\mu$. Inhomogeneous
fluctuations of $\psi_\mu$ could also be the source of the
Q-component of energy~\cite{Caldwell+Dave+Steinhardt:1998}.

\section*{Acknowledgments}

JWM is supported by the Natural Sciences and Engineering Research
Council of Canada.


\end{document}